\documentclass[showpacs,preprintnumbers,amsmath,amssymb,prl,twocolumn]{revtex4}
\def\nablab{{\mbox{\boldmath $\nabla$}}}

\usepackage{graphicx}
\usepackage{dcolumn}
\usepackage{bm}

\begin{document}

\title{
Duality and self-duality in Ginzburg-Landau theory
with Chern-Simons term
}
\author{Hagen Kleinert}
\email{kleinert@physik.fu-berlin.de}
\homepage{http://www.physik.fu-berlin.de/~kleinert/}
\author{Flavio S. Nogueira}
\email{nogueira@physik.fu-berlin.de}
\affiliation{Institut f\"ur Theoretische Physik,
Freie Universit\"at Berlin, Arnimallee 14, D-14195 Berlin, Germany}

\date{Received \today}

\begin{abstract}
We derive the exact dual theory
of  a lattice Ginzburg-Landau theory
with an additional topological
Chern-Simons (CS) term.
It is shown that in the zero-temperature limit, the statistical
parameter $\theta=1/2\pi$ corresponds to a fixed point of the
duality transformation. Thus we have found  a nontrivial example of
self-duality in three dimensions. In this scenario, the
specialization of anyonic to fermionic statistics my be viewed as
a phase transition.
\end{abstract}

\pacs{74.20.-z, 05.10Cc, 11.25.Hf}
\maketitle

Duality is an important tool for studying the strong-coupling regime
of field theories.  Most powerful is the property of self-duality,
which in historic example of the two-dimensional Ising model
 allowed for an exact determination of the critical
temperature before the Onsager solution \cite{Kramers}.  A direct
consequence of self-duality is the exact
equality of the amplitudes of
the leading term $A_\pm|T/T_c-1|$  of the specific heat above and below the critical
temperature: $A_+/A_-=1$.  The same result holds for two-dimensional
$q$-state Potts model \cite{Kaufman} since it is also self-dual.

There are only very few examples
for self-duality in dimension $d>2$
\cite{Itzykson}.
The four-dimensional $Z_2$ gauge theory is self-dual determining
exactly
 the critical temperature.
 In three dimensions, the $Z_2$ gauge theory
coupled to bosonic matter fields is self-dual. A
common feature of these examples is the discreteness of the
symmetry group. For continuous symmetries, non-trivial models
with self-duality are very hard to find.

Take for example
the three-dimensional
$U(1)$ theory
discussed by  Townsend {\it et al.} \cite{Townsend}
and Deser and Jackiw
\cite{Deser}.
It is self-dual, but noninteracting,
  describing a vector field ${\bf A}$
with a Hamiltonian \cite{Townsend,Note}

\begin{equation}
\label{Townsd}
{\cal H}_0=\frac{\theta^2}{2}{\bf A}^2+i\frac{\theta}{2}{\bf A}\cdot(
\nablab\times{\bf A})
\end{equation}
Its self-duality is obvious by the equation of
motion ${\bf A}=-i(\nablab\times{\bf A})/\theta$, and
$\nablab\times{\bf A}$ is  dual to ${\bf A}$. It has been pointed
out by Deser and Jackiw \cite{Deser} that Eq. (\ref{Townsd}) is
equivalent to the locally gauge-invariant model

\begin{equation}
\label{freecs}
{\cal H}_0'=\frac{1}{2}(\nablab\times{\bf B})^2+i\frac{\theta}{2}
{\bf B}\cdot(\nablab\times{\bf B}).
\end{equation}
The Hamiltonian (\ref{Townsd}) is in fact the dual of the
Hamiltonian (\ref{freecs}), the latter being
locally gauge-invariant, while the former is not.

In this note we want to exhibit a nontrivial
self-dual model
containing a topological Chern-Simons (CS) term \cite{Jackiw}
 which may serve as an
effective field theories
for condensed matter systems.
Such effective field theories
have recently become
useful tools for understanding
a variety of new low-temperature phenomena \cite{Fradkin}.
A CS field permits us to attach flux tubes to particles
in three dimensions, thereby  changing
continuously their statistics from bosonic to fermionic
\cite{Wilczek}. In the context of the fractional quantum Hall
effect, the CS mass is associated to the phase determining the
statistics of Laughlin quasi-particles \cite{Fradkin}. A typical
effective theory in this context consists of a Ginzburg-Landau (GL)
the the Maxwell term $(\nablab\times {\bf A})^2/2$ is exchanged
by a CS-term \cite{Zhang}. If we want to perform
renormalization studies of such a model,
the Maxwell term must be included in addition
to the CS term
to provide
the theory with a gauge-invariant cutoff \cite{Semenoff}.
This model will be  be called
GLCS model,
 and has a Hamiltonian

\begin{eqnarray}
\label{model}
{\cal H}_{e}&=&\frac{1}{2e^2}(\nablab\times{\bf A})^2+
i\frac{\theta}{2}{\bf A}\cdot(\nablab\times{\bf A})+
|(\nablab-i{\bf A})\psi|^2\nonumber\\
&+&r|\psi|^2+\frac{u}{2}|\psi|^4,
\end{eqnarray}
where $\psi$ is a complex order field.
The limiting Hamiltonian ${\cal H}_\infty$ gives an effective
description of anyonic quasi-particles for
statistical parameters $0<\theta<1/2\pi$. A bosonized fermion theory
lies at the upper end of this interval: $\theta=1/2\pi$.
For arbitrary $e$ and $\theta$,
the critical behavior of the above model
has  been studied before
\cite{KleinertCS}. It exhibits continuously varying critical exponents
 due to the fact
that the CS-term remains unrenormalized \cite{Semenoff}.
More recently, the GLCS model appeared as an effective dual theory
for the spin sector of a strongly correlated electron system
\cite{Nayak}. The spin sector is originally a fermionic theory,
describing part of the effective dynamics of the $t-J$ model,
which in bosonized form contains a CS-term. Since the
low-energy theories associated with the
 $t-J$ model
are typically gauge theories at infinite
coupling, the corresponding bosonized theory has $\theta=1/2\pi$ and
no Maxwell term. The corresponding dual theory of
the spin sector, however, contains a Maxwell term and has the
form
 (\ref{model}) \cite{Nayak}. This is to be expected, since
a duality transformation maps a strongly coupled to a weak
coupled  theory.

Here we shall discuss the interesting
 duality properties of the Hamiltonian
${\cal H}_\infty$ using the approach introduced for the GL model
in Refs. \onlinecite{Kleinert} and \onlinecite{KleinertBook}.
We perform the duality transformation {\it exactly} in a
lattice GLCS model at $e=\infty$ and show that this
generates a Maxwell term in the dual theory.
The duality transformation will be
shown to have a fixed point at $\theta=1/2\pi$ and
zero temperature. There the theory is self-dual.
At the parameter
$\theta=1/2\pi$ the statistics of the
order field becomes
purely fermionic.
The self duality extends over the entire
critical regime
of the model since
the generated Maxwell
is irrelevant for the
renormalization group flow.
Our results implies
that
anyons become fermions in a phase transition.

The duality transformation of ${\cal H}_\infty$ can be
performed {\it exactly}
if we consider the London limit
in which the size of the complex order field is fixed, put the
model on a lattice, and approximate it \`a la Villain
with a high accuracy \cite{JK}.
Within this philosophy, the lattice version
of the Hamiltonian ${\cal H}_e$ has the
 form
\begin{eqnarray}
\label{Lmodel}
H_{e}&=&\sum_x\left[\frac{\beta J}{2}\sum_{\mu}(\nablab_\mu\varphi_x-2\pi
n_{x\mu}
 -A_{x\mu})^2\right.\nonumber\\
&+&\left.\frac{1}{2e^2}(\nablab\times{\bf A}_x)^2
+i\frac{\theta}{2}{\bf A}_x\cdot(\nablab\times{\bf A}_x)\right],
\label{@thenform}\end{eqnarray}
where the lattice derivative is
$\nablab_\mu f_x\equiv f_{x+\mu}-f_x$, and $\beta=1/T$.
The partition function
is given by

\begin{equation}
\label{partition}
Z=\int\left[\prod_x \frac{d\varphi_x}{2\pi} d{\bf A}_x\right]\sum_{{\bf n}_x}
\exp(-H_e),
\end{equation}
where the sum runs over all integer
$n_{x\mu}$ and  the domains of integration are $\varphi_x\in(-\pi,\pi)$
and $A_{x\mu}\in(-\infty,\infty)$.

Let us set $e^2=\infty $.
Following standard techniques
\cite{Kleinert,KleinertBook,Peskin,Dasgupta}
we rewrite
$({\beta J}/{2})(\nablab_\mu\varphi_x-2\pi n_{x\mu}
-A_{x\mu})^2$ as
${\bf B}^{2}_x/(2 \beta J) +i{\bf B}_{x\mu}(\nablab_\mu\varphi_x-2\pi n_{x\mu}
-A_{x\mu})$ and apply Poisson's  formula  to convert the
integrals over ${\bf B}_x$ into a sum over integer variables ${\bf b}_x$,
we
 obtain the dually transformed
Hamiltonian:
\begin{equation}
\label{H1}
H'_\infty=\sum_x\left[\frac{1}{2\beta J}{\bf b}_x^2+i{\bf b}_x\cdot
{\bf A}_x+i\frac{\theta}{2}{\bf A}_x\cdot(\nablab\times{\bf A}_x)\right],
\end{equation}
where the integer variables ${\bf b}_x$
 satisfy the
constraint $\nablab\cdot{\bf b}_x=0$
arising from the $\varphi_x$ integration, meaning that only
configurations with closed vortex loops are counted
\cite{KleinertBook}. We
introduce a second integer variable $\tilde{\bf a}_x$ such
that ${\bf b}_x=\nablab\times\tilde{\bf a}_x$. Using the Poisson
formula to convert the sum over $\tilde{\bf a}_x$ to
an integral over $\tilde{\bf A}_x$ and an auxiliary
sum over integers $\tilde{\bf b}_x$ yields
\begin{eqnarray}
\label{H2}
H''_\infty&=&\sum_x\left[\frac{1}{2\beta J}(\nablab\times\tilde{\bf A}_x)^2+i{\bf A}_x
\cdot(\nablab\times\tilde{\bf A}_x)\right.\nonumber\\
&+&\left.i\frac{\theta}{2}{\bf A}_x\cdot
(\nablab\times{\bf A}_x)+2\pi i\tilde{\bf b}_x\cdot\tilde{\bf A}_x\right],
\end{eqnarray}
where $\nablab\cdot\tilde{\bf b}_x=0$. Integrating out ${\bf A}_x$, we
obtain
\begin{eqnarray}
\tilde{H}_\infty&=&\sum_x\left[\frac{1}{2\beta J}
(\nablab\times\tilde{\bf A}_x)^2+i\frac{1}{2\theta}\tilde{\bf A}_x\cdot
(\nablab\times\tilde{\bf A}_x)
 \right.
\nonumber \\&&
\left.~~~~~~+i~2\pi \tilde{\bf b}_x\cdot\tilde{\bf A}_x+
\frac{ \epsilon _0}{2}{\tilde{\bf b}}_x^2\right],
\label{H3}
\end{eqnarray}
where we have added an extra  core energy
to the $\tilde{\bf b}_x$ field, thereby generalizing slightly the model
which has $ \epsilon _0=0$.
The core energy allows us
to introduce an auxiliary field
$\tilde\varphi_x$ and
rewrite  the partition function
as

\begin{eqnarray}
\label{Hdual}
\!\!\!\!H_\infty^{\rm dual}\!\!\!&=\!\!&
\sum_x\left[
\sum_{\mu}\frac{1}{2\epsilon_0}\left(\nablab_\mu\tilde\varphi_x
-2\pi \tilde {\bf n}_{x}-\tilde{\bf A}_x\right)^2\nonumber\right. \\\!\!\!&+\!\!&\left.
\frac{1}{8\pi^2\beta J}
(\nablab\times\tilde{\bf A}_x)^2\!+i\frac{1}{8\pi^2\theta}
\tilde{\bf A}_x\cdot(\nablab\times\tilde{\bf A}_x)\right]\!,
\end{eqnarray}
where we have rescaled $\tilde{\bf A}_x\to\tilde{\bf A}_x/2\pi$.
The first term
arises from $ { \epsilon _0}\tilde{\bf b}_x^2/2+
i~2\pi \tilde{\bf b}_x\cdot\tilde{\bf A}_x
$ in Eq.~(\ref{H3})
by the same standard
technique described before  Eq.~(\ref{H1})
for the same untilded quantities,
although in the opposite direction.
The dual Hamiltonian (\ref{Hdual}) has precisely
 the same form as
(\ref{@thenform}), and
coincides with it
 if we replace
 $\epsilon_0\rightarrow (\beta J)^{-1}$, $e^2\rightarrow 4\pi^2\beta J$,
$\theta\rightarrow 1/(4\pi^2\theta)$.
The dual Hamiltonian
is the limit $ \epsilon _0\rightarrow 0$ of (\ref{H3}),
in which case we speak with Peskin \cite{Peskin}
of a ``frozen'' limit of the GLCS model. Thus, up to smooth factors in the
temperature, the free energy satisfy the duality relation:

\begin{equation}
\label{dualrel}
F(T,\theta,e=\infty)=F\left(T'=0,~\theta'=\frac{1}{4\pi^2\theta},
~e'=2\pi\sqrt{\frac{J}{T}}\right).
\end{equation}
From Eq. (\ref{dualrel}) we see that the fixed point of the duality
transformation is given by

\begin{equation}
\label{fixpdual}
T=0,~~~~~~\,\theta'=\theta=\frac{1}{2\pi}.
\end{equation}
At this point,
the Hamiltonian $H_\infty$
is self-dual.
Note that in the absence of the CS-term,
the dual of the frozen superconductor is merely a Villain model
as observed by Peskin \cite{Peskin}, and there is no self-duality.

The Villain model can well be approximated by a $XY$-model
which, in turn, can be transformed into a complex disorder field
theory \cite{KleinertBook}.
 The result is
\begin{eqnarray}
\label{disfth1}
{\cal H}_\infty^{\rm dual}&=&\frac{1}{8\pi^2g^2}(\nablab\times\tilde{\bf A})^2
+i\frac{1}{8\pi^2\theta}\tilde{\bf A}\cdot(\nablab\times\tilde{\bf A})\nonumber\\
&+&\left|\left(\nablab-i\tilde{\bf A}\right)\phi\right|^2
+r'|\phi|^2+\frac{u'}{2}|\phi|^4,
\end{eqnarray}
where $g^2\equiv\beta J \Lambda$, and $\Lambda$ is an ultraviolet
cutoff reminiscent of the lattice.
Interestingly, the local gauge invariance
of the disordered phase of ${\cal H}_\infty$ is also present in the
disordered phase of ${\cal H}_\infty^{\rm dual}$. This does not happen in
the GL model, where the disordered phase of the corresponding
dual theory has no local gauge symmetry \cite{Kleinert,KleinertBook,Kiometzis}.
Note that if we set $\theta=1/2\pi$ in Eq. (\ref{disfth1}), we obtain
a continuum dual Hamiltonian similar to the one considered in
Ref. \onlinecite{Nayak}, except that the coefficient of the
Maxwell term is not equal to unity. The continuum dual Hamiltonian
(\ref{disfth1}) was obtained for the first time in the context of
the fractional quantum Hall effect by Wen and Niu \cite{Wen}.
Their derivation, however, was only heuristic and performed in
the continuum, in contrast to our
derivation by an  {\it exact} duality transformation
on the lattice.

Near the critical region, where the lattice model
has a good continuum limit,
the Maxwell term is irrelevant with respect to the CS-term
\cite{Semenoff}, which tells us that
the
continuum dual model (\ref{disfth1}) becomes
 self-dual
for
$\theta=1/2\pi$. {\it This observation
implies that an
interacting fermion model in three dimensions corresponds to
the fixed point of a duality transformation in the} GLCS model.
This self-dual point marks  a phase
transition in the GLCS model as the statistical parameter $\theta$ is
varied. This phase transition is associated
with the specialization
of
anyonic statistics to fermionic statistics.

In the original Hamiltonian
${\cal H}_\infty$, the $\beta$-function
$\beta_{\theta_r}\equiv\mu\partial\theta_r/\partial\mu=0$
\cite{Semenoff,Chen},
with $\theta_r$ being the renormalized statistical parameter.
The $\beta$-function vanishes because
the anomalous dimension
of the
the vector potential vanishes, $\eta_A=0$. This is in
contrast to the pure GL model, where $\eta_A=1$ \cite{Herbut}.
In our model,
 the scaling dimension of ${\bf A}$ is
equal to the canonical one, $[{\bf A}]=1$. This
makes the scaling dimension of the Maxwell term equal to four,
and thus irrelevant
in the long-wavelength limit.

In the
dual model,  it is the $\beta$-function of the ratio $\hat{g}_r^2/\theta_r$
that vanishes at the critical point. Here, $\hat{g}_r^2=g_r^2/\mu$ is the renormalized
dimensionless gauge coupling of
the disorder field theory. Thus, the $\beta$-functions of
$\hat{g}_r^2$ and $\theta_r$ in the disorder field theory
are given by

\begin{equation}
\label{betafs}
\beta_{\hat{g}^2}=(\eta_{\tilde A}-1)\hat{g}_r^2,~~~~~~~~~~~\beta_{\theta_r}=
(\eta_{\tilde A}-1)\theta_r,
\end{equation}
where $\eta_{\tilde A}\equiv\mu\partial\ln Z_{\tilde A}/\partial\mu$, with $Z_{\tilde A}$ being the
wave function renormalization of the dual gauge field.
From (\ref{betafs}) we see that a charged fixed point
corresponds to $\eta_{\tilde A}=1$, in which case $\theta_r$ can assume
any value. Due to the presence of the Maxwell term, the
canonical dimension of $\tilde{\bf A}$ is two and since
$\eta_{\tilde A}=1$, the scaling dimension of $\tilde{\bf A}$ is unity. 
Thus, we
see that ${\bf A}$ and its dual $\tilde{\bf A}$ have the same
scaling dimension, that is, $[{\bf A}]=[\tilde{\bf A}]=1$, and we can
speak of a self-duality of the gauge field scaling dimension. In the
GL model, the scaling dimensions behave differently under duality.
Indeed, in the GL model, we have $\eta_A=1$
\cite{Herbut}, whereas  the corresponding
disorder field theory has $\eta_{\tilde A}=0$ \cite{Kiometzis,deCalan,Hove}
and, since the Maxwell term is
present in both theories, we have $[{\bf A}]=1$ and
$[\tilde{\bf A}]=2$.

Summarizing, we have found a nontrivial self-dual model
in three dimensions as a fixed point in an
 exact duality transformation of a
lattice model containing
 a scalar model minimally coupled
to a gauge field, whose dynamics is governed by a CS
term. The lattice dual model contains a Maxwell term that vanishes
at the self-dual point. This lies at the anyonic
statistics parameter
 $\theta_c=1/2\pi$,  corresponding to Fermi
statistics. The self-duality implies that there exists a
phase transition if the statistics parameter is varied.
We have also derived the continuum limit of the dual model
as  a disorder field theory. It was shown that
the original gauge field and its dual have the same scaling dimension.
An interesting topic to study in the future concerns the order of
the statistics-parameter-induced phase transition.

\begin{acknowledgments}
The work of FSN is supported by the Alexander von Humboldt foundation.
\end{acknowledgments}


\begin{thebibliography}{100}
\bibitem{Note} In this paper the term ``Hamiltonian'' is used
to mean ``Euclidian Lagrangian''.
\end{thebibliography}
\end{document}